\def\beg{\begin{equation}}
\def\eeq{\end{equation}}
\documentstyle[12pt]{article}
\begin{document}
\begin{center}
{\Large{\bf Comments on ``Fractional Quantum Hall Effect of Composite Fermions" by W. Pan et al Phys. Rev. Lett. {\bf 90}, 016801 (2003) }}
\vskip0.35cm
{\bf Keshav N. Shrivastava}
\vskip0.25cm
{\it School of Physics, University of Hyderabad,\\
Hyderabad  500046, India}
\end{center}

It is shown that even number of flux quanta are not attached to one electron. The magnetic flux is not detached from the currents and 
the {\bf E} and {\bf H} separation does not occur in the quantum 
Hall effect, where {\bf E} is the electric vector and {\bf H} is 
the magnetic vector of the electromagnetic field. We show how a 
series can agree with the experimental data and be wrong also. It 
is not possible for three electrons to carry 8 flux quanta. There 
is no temperature in the composite fermion formulas. It may be 
perfectly acceptable to modify the Biot and Savarts's law which 
is the fundamental law used for making electromagnets but the
 CFs are internally inconsistent and hence are not suitable as
 part of fundamentally correct physics.
\vfill
Corresponding author: keshav@mailaps.org\\
Fax: +91-402-301 0145.Phone: 301 0811.
\newpage
\baselineskip22pt
\noindent {\bf 1.~ Introduction}

     Recently, Pan et al[1] have discussed the applicability of the composite fermion (CF) model to the actual experimental data of 
quantum Hall effect. It is not clear whether this model is 
applicable to the data or not. Therefore, we shall make an effort 
to analyse the applicability of the CF model to the real experimental measurements. Our conclusion from this study is that the CF model 
is internally inconsistent and should be discarded. The cases, 
where there is an apparent agreement between the model and the data, 
are a  result of misappropriations.

\noindent{\bf 2.~~Comments.}

{\it (i) Odd denominator with out mechanics }

     Pan et al suggest that CF model has been very successful in providing a rationale for the observed sequences at the Landau 
level filling factors, $\nu=p/(2mp\pm 1)$, where $p$ is an integer 
and $m$ is also an integer. These integers are not derived from any mechanics but will be adjusted to give experimental values. It is 
true that both of these series are the same as found in the 
experimental measurements particularly when tabulated for various 
values of $p$ and $m$. So when the series are correct, then why the model should not be?  Let us say that the model is correct and beat drums about it. So can a wrong
 model become correct because there is sound of drums. No. The model 
is wrong. In the begining of the subject, it was found that odd denominators occured. So let us construct the odd number. Take an 
even number and 
add $\pm 1$ so the odd number is correct. That means the generated 
odd number will agree with the data. Now, look at the numerator of 
the experimental data and set it in the formula so that the numerator 
as well as the denominator give the experimental value. In this way 
the CF model sequence becomes the same as the experimental. Now, we start saying that the model has great success without deriving the series from the Schrodinger equation. So the CF series which has great
success does not have a theoretical basis.

     Usually, a correct theory is derived from quantum mechanics but 
the CF sequences which give the correct experimental values have not been derived by using quantum mechanics. Some times, a model such as Ginsburg-Landau model gives several features which agree with the 
data but CF is not based on Ginsburg-Landau model or any other model. 
So it is a new model, but then it should be internally consistent. We will see that the CF model is neither internally consistent, nor agrees with the data. For example, when two flux quanta are attached to the electron, it will be very heavy. The mass of a CF will be several hundred times the mass of the electron. Similarly, the size of the CF 
is much larger than that of the electron but the CF formula uses same density for CFs as for the electrons. Therefore, the CF model is internally inconsistent as far as density is concerned. Therefore, the model which agrees with the data is not necessarily a correct model. 
The sequence is correct but the flux attachment is wrong and masses of CF are not in agreement with data.

{\it(ii) Independent particles.}

     Pan et al suggest that CFs are independent particles. Usually,
the flux is produced when a current passes in a coil. The current is 
the flow of electrons. Therefore, flux depends on the flow of 
electrons and flux is not a particle but it arises due to electrons.
Therefore, flux is not an independent particle. The Biot and Savart's
law in the classical electrodynamics describes the field produced by flow of current in a coil. Therefore, flux depends on electrons and hence flux attached to electrons is impossible. Let us detach the 
flux from the current. That means {\bf H} will be detached and 
electrons are left with {\bf E} only. The {\bf H} detached from one electron is to be attached to another electron. More flux has to be  detached from another electron and attached to the first electron 
untill two flux quanta are attached to the first electron. This is called CF and this is impossible. Neither {\bf H} decouples from 
{\bf E} nor flux can be attached to an electron.  For the CF model, 
we have to attach two flux quanta to one electron. In this case, we 
have to have three electrons, two of which have detached their field which has been attached to one of them. The CF model attaches two 
flux quanta to one electron but does not mention the two electrons 
which have lost their flux. Therefore, CF model is inconsistent. In 
the comment (i) we have mentioned that the series of fractions is correct but it is obvious that CF model does not detach flux and hence is inconsistent because it adds flux without detaching. That means CF can create flux from nothing. Further, Pan et al's suggestion that CFs are independent particles, is not correct because there are currents.

{\it(iii) Density.}

     The CF magnetic field is an effective field which is reduced 
due to attachment of flux quanta,$B*=B-2np\phi_o$ where $n$ is the density of electrons per unit area and $p$ is an integer. This field does not depend on the number density of CFs. If both CFs and 
electrons are present in the system then the field should use both 
the densities.

{\it(iv) Sequences.}

     Pan et al suggest that "Prime examples for the applicability 
of the CF model are the sequences at $\nu=p/(2p\pm 1)$". First of all the sequence can agree with the data only when it has been derived 
from some physics. Since the sequence has been obtained from the 
data itself then there is no question of comparing the sequence  
with the data. An object A can agree with the object B when A is 
derived from a theory and B from the data. If A has been obtained 
from the data and B is the data, then A and B are one and the 
question of agreement of A with B does not arise.

{\it(v) Conditions at hand.}

   Page 016801-1, left column, para 2: `` ... exact applicability to 
the conditions at hand is doubtful". The attachment of flux quanta 
to the electrons is surely incorrect. The filling factors of 4/11, 
5/23, 7/11, 4/13, 6/17 and 5/17 are obtained and it is claimed that these are due to interactions. The 4/11 is found to be spin polarized and 1-4/11 = 7/11 is found to rapidly disappear and its spin polarization is uncertain. As far as CF model is concerned, the sequences which agree with the data are ``independent of spin". Therefore, the CF are not spin polarized. In fact the CF model is completely independent of spin. So no spin effects can be associated with CF. The CF is spinless so can not belong to real physical situation. There is no prescription in the CF model to incorporate interactions due to lack of mechanics. 

{\it(vi) Correction.}

Page 016801-2, right column, 2nd para: ``4/11 is created by the  following mental sequence ... every three electrons carry 8 flux quanta". This is not physics.
The CF model can provide 6 flux quanta to three electrons but not 8
but the model does not tell where these flux quanta come from.
Most of the material on page 016801-2 is not based on any hamiltonian and hence is purely "hypothetical" and not derived from proper logic. Any way, there is no way 8 flux quanta could be attached to three electrons. The suggested correction  is not provided for in the CF model. 

{\it (vii) Temperature.}

     Page 016801-3, left column, 2nd para: ``The enormous change of shape of the data around $\nu=7/11$
leads to erratic temperature dependence". The CF model does not provide any prescription to study temperature dependence but the experimental data varies with temperature.

{\it(viii) Spin.}

     Page 016801-3 right column, 2nd para: ``We therefore refrain from assigning any spin polarization to $\nu$ =7/11 state". The CF model does not have the capability to assign spin.

{\it (ix) Stability.}

     Page 016801-3, right column, 3rd para: `` ... theoretical calculations as to the stability of such higher order FQHE states are difficult...". When even in the case of lower order, the flux quanta are not attached to the electrons, there is no reason to make effort to understand higher order states using flux-quanta attachment model.
The CF model does not have a theoretical base so it is not possible to calculate stability of states in any order.

{\it (x) Incompressibility}

     Page 016801-4, left column, 2nd para: `` Numerical calculations seem to exclude the existence of incompressible states at such filling factors for spin polarized systems, at least for zero-thickness systems".  The incompressibility is associated with Laughlin wave function. However, Laughlin was not able to resolve the area from the charge. There is no mention of compressibility in the CF model.

\noindent{\bf3. ~~ Additional Comments}

     It has been shown elsewhere[2] that mass of the CF is several thousand times the mass of the electron whereas the experimental mass 
is $0.4m_e$. The CF model works because of too many parameters[3]. 
The flux attaching and detaching violates classical electrodynamics. The flux detaching has not been considered in the CF model which makes it inconsistent[4]. The CF attaches flux on one electron without detaching from another and hence it is inconsistent. The CFs are found to be too big in size to fit the given space [5]. It is found that Laughlin's exact calculation was not able to distinguish between area and the charge and hence can not belong to real systems[6].

\noindent{\bf4.~~ Conclusions}.

     Page 016801-4, right hand column, top line: `` Such a self-similarity in the sequence of FQHE states is too appealing to be discarded yet".

     There is no theoretical basis for CF and the requirement of flux attachment is not satisfied in real systems. The sequence $\nu=p/(2mp\pm 1)$ has neither been derived from quantum mechanics nor from classical mechanics. It is not derivable from Maxwell equations of classical electrodynamics. The CF model is internally inconsistent. The sequence agrees with the data because it is derived from the data itself. Hence, there is no agreement between the CF model and the data. Under these circumstances, the CF model will have to be discarded. Lack of theoretical basis in the CF model has also been noted by Dyakonov[7] and Farid[8].

\noindent{\bf About the author}: {\it Keshav Shrivastava has obtained Ph.D. degree from the
Indian Institute of Technology and D. Sc. from Calcutta University. 
He is a member of the American Physical Society, Fellow of the Institute of Physics (U.K.)and Fellow of the National Academy of Sciences, India. He has worked in the Harvard University, University of California at Santa Barbara, the University of Houston and the Royal Institute of Technology Stockholm. He has published 170 papers in the last 40 years. He is the author of two books}.

The correct theory of quantum Hall effect is given in ref.9.
\vskip1.25cm

\noindent{\bf5.~~References}
\begin{enumerate}
\item W. Pan, et al, Phys. Rev. Lett. {\bf 90}, 016801 (2003)
\item K. N. Shrivastava, cond-mat/0211223.
\item K. N. Shrivastava, cond-mat/0210320.
\item K. N. Shrivastava, cond-mat/0209666.
\item K. N. Shrivastava, cond-mat/0209057.
\item K. N. Shrivastava, cond-mat/0212552.
\item M. I. Dyakonov, Talk at NATO ARW ``Recent trends in theory 
of physical phenomena in high magnetic fields", cond-mat/0209206.
\item B. Farid, cond-mat/0003064.
\item K.N. Shrivastava, Introduction to quantum Hall effect,\\ 
      Nova Science Pub. Inc., N. Y. (2002).
\end{enumerate}
\vskip0.1cm
Note: Ref. 9 is available from:\\
 Nova Science Publishers, Inc.,\\
400 Oser Avenue, Suite 1600,\\
 Hauppauge, N. Y.. 11788-3619,\\
Tel.(631)-231-7269, Fax: (631)-231-8175,\\
 ISBN 1-59033-419-1 US$\$69$.\\
E-mail: novascience@Earthlink.net

\vskip5.5cm
\end{document}